\documentclass[10pt]{iopart}
\usepackage{iopams,bm,cite,color,graphicx,mathptmx,url,hyperref}

\begin{document}

\title{The many facets of the Fabry-Perot}

\author{Luis~L~S\'{a}nchez-Soto$^{1,2}$,  Juan~J~Monz\'on$^{1}$ and Gerd
Leuchs$^{2,3}$}

\address{$^{1}$ Departamento de \'Optica, Facultad de F\'{\i}sica,
  Universidad Complutense, E-28040~Madrid, Spain}
 
\address{$^{2}$ Max-Planck-Institut f\"ur die  Physik des Lichts,  
G\"{u}nther-Scharowsky-Stra{\ss}e 1, Bau 24,  D-91058 Erlangen, Germany}

\address{$^{3}$ Institut f\"ur Optik, Information und Photonik,
 Universit\"{a}t Erlangen-N\"{u}rnberg,  Staudtstra{\ss}e~7/B2,
 91058 Erlangen, Germany}

\date{\today}

\begin{abstract}
  We address the response, both in amplitude and intensity, of a
  Fabry-Perot from a variety of viewpoints. These complementary
  pictures conspire to achieve a comprehensive and consistent theory
  of the operation of this system.
\end{abstract}

\noindent {\it Key words:} Fabry-Perot, Interference. 
 
\eqnobysec

%\maketitle

\section{Introduction}

The names of Charles Fabry and Alfred Perot\footnote{Is it Perot or
  P\'erot?  The former accent-free spelling is indeed the official
  one, as confirmed by Perot's birth certificate and other official
  documents~\cite{Georgelin:1995aa,Metivier:2006aa}. This is also the
  form used by the most authoritative French authors, such as
  Kastler~\cite{Kastler:1974aa}, Fran\c{c}on~\cite{Francon:1966aa}
  Jacquinot~\cite{Jacquinot:1960aa}, Chabbal~\cite{Jacquinot:1956aa}
  or Connes~\cite{Connes:1986aa}, among others. We therefore adopt here this
  usage. However, the accented version persists to this day in the
  scientific literature. Surprisingly enough, Perot referred to
  himself as P\'erot in a few of his original works. In an awesome
  paper~\cite{Steel:1999aa}, Steel hypothesises that this
  misspelling had its origins in Parisian journals, such as
  Comptes Rendues, the savant editors of which might have
  over-zealously purported to know better than authors from the south
  of France. However, as Orr~\cite{Orr:2006aa} aptly points out, it seems
  that the error became entrenched on the other side of the Atlantic,
  where Perot's name was persistently misspelt in American digests of
  his papers during the period 1900--1905.}  are inextricably linked
to one of the most deceptively simple setups ever devised in
optics:  just two parallel highly reflecting
mirrors. 

The instrument was the fortunate convergence of two independent
developments: a long history of producing mirrors and a deeper
understanding of multiple beam interference~\cite{Connes:1986aa}.
Fabry and Perot took full advantage of the potential of this setup and
accomplished major scientific discoveries: some of their seminal work
includes the determination of the temperature of the Orion nebula and
the measurement of the gravitational redshift of light.  Other
discoveries are less known: the determination of the
altitude and thicknesses of the atmospheric ozone layers; the
calibration of the flux of different stars, the improvement of stellar
spectrophotometry; the development of electrometers to measure weak
potentials; the elaboration of an atlas of emission lines; the
laboratory verification of the Doppler-Fizeau principle and much
more~\cite{Amram:2000aa}. A detailed biographic sketch of these two
influential scientists can be found in the excellent article by
Mulligan~\cite{Mulligan:1998aa}.

Apart from its luminosity, the distinctive feature of the Fabry-Perot
is its narrow resonances. This is the basis for its extensive use in 
high-resolution spectroscopy, interferometry, and laser resonators. 
An exhaustive covering can be found in the two thorough monographs by
Hernandez~\cite{Hernandez:1986zr} and Vaughan~\cite{Vaughan:1989qn}.
This variety of contexts accounts for the diversity of terms used to
name it: interferometer, spectrometer, filter, \'etalon and cavity, to cite
only the most popular.

Each of these tags capitalizes on specific ideas.  The geometric
treatment, in which one adds the multiple beams reflected at each of
the different interfaces, is probably the more instructive and,
accordingly, is reproduced in almost every
textbook~\cite{Born:1999yq}.  But the question can also be tackled
from different standpoints. In the course of many years of teaching
and research, we have brought together a number of nonconventional
approaches to the Fabry-Perot. The delightful notes compiled by
Jacobs~\cite{Jacobs:sy} persuaded us that it is worthwhile to
understand this remarkable device in as many different ways as
possible. It is not that the classical textbooks are incomplete;
rather, we seek to highlight some points that often go unnoticed and
have been elevated to a position of relevance by current
progress. Besides its inherent interest, we believe that this can help
overcome the little cross talk between different specialists.

A final word of caution: the arrangement of a realistic Fabry-Perot
imposes stringent practical requirements that the interested reader
can find in the two aforementioned treatises. Our discussion, however,
is limited to its basic aspects. None of the methods presented here is
original by itself and we make no claim of completeness of the
collection. Although, at first sight it seems difficult to find
anything truly new and useful in this topic, we hope that the final
picture emerging from all these complementary viewpoints gives fresh
perspectives of the subtleties behind this amazing system.

\section{Amplitude response of a Fabry-Perot}

\subsection{Multiple beam interference}
\label{sec:mbi}

The ideal Fabry-Perot (FP) interferometer consists of two parallel
mirrors (which, for simplicity, we assume to be identical) separated
at a distance $d$. This is customarily modeled by a plane parallel
plate of thickness $d$ and refractive index $n$ immersed in a medium
of index $n^{\prime}$.  The plate is illuminated near normal incidence
with a linearly polarized quasi-monochromatic plane wave, with the
electric field lying either parallel or perpendicular to the plane of
incidence. Any diffraction effect or polarization dependence are thus
neglected henceforth.

\begin{figure}[b]
 \centering{\includegraphics[height=4.5cm]{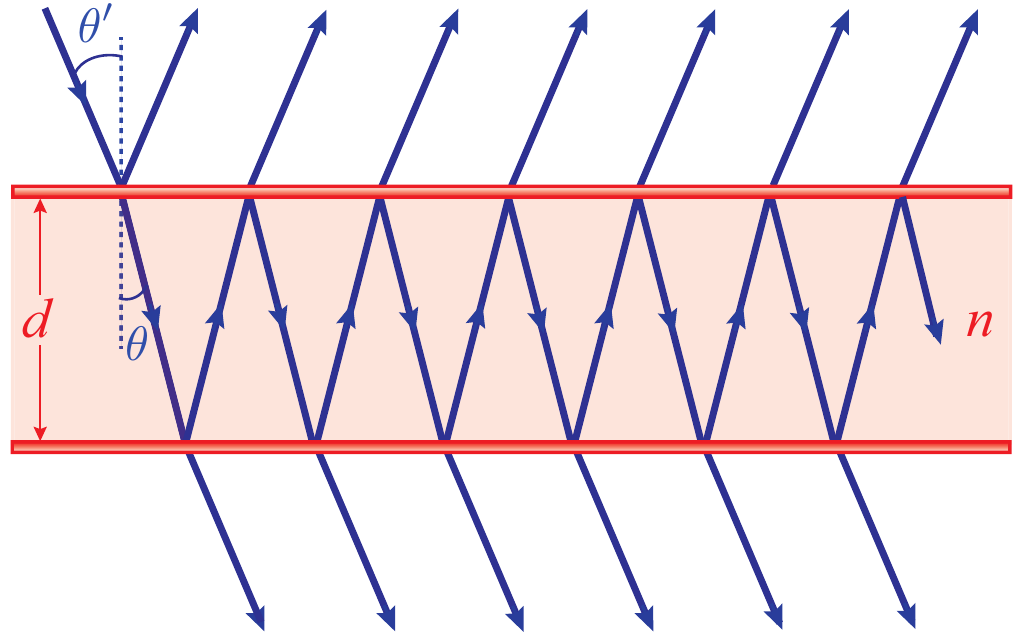}}
  \caption{Schematic of the reflected and transmitted amplitudes in an
    FP, modeled as a plane parallel plate.}
  \label{fig:schema}
\end{figure}

We denote by $A_{\mathrm{in}}$ the amplitude of the incident wave. At
the first surface, this wave is divided into two plane waves, one
reflected and the other transmitted into the plate. The latter wave is
incident on the second surface and is divided into two plane waves,
and the process of division of the wave remaining inside the plate
continues as sketched in figure~\ref{fig:schema}.

%%%%%%%%%%%%%%%%%%%%%%%%%%%%%%%%%%%%%%%%%%%%%%%%%
%%%%%%%%%%%%%%%%%%%%%%%%%%%%%%%%%%%%%%%%%%%%%%%%

Let $r^{\prime}$ and $t^{\prime}$ be the Fresnel reflection and transmission
coefficients, respectively, for a wave travelling from the surrounding
medium into the plate and let $r$ and $t$ the
corresponding coefficients for a wave travelling from the plate to
the surrounding medium.  The total reflected and transmitted waves can
be jotted down as
\begin{eqnarray}
  \label{eq:TampFP1}
A_{\mathrm{ref}} ( \varphi ) = A_{\mathrm{in}}   [ r^\prime + 
tt^\prime  r \rme^{-\rmi 2 \varphi}   ( 1 +  {r}^{2} \rme^{- \rmi 2 \varphi} + 
  {r}^{4} \rme^{- \rmi 4  \varphi}  + \ldots ) ] \, , \nonumber  \\
 \\
  A_{\mathrm{trans}} ( \varphi ) = A_{\mathrm{in}}  \,  tt^{\prime}
  \,  \rme^{- \rmi \varphi} 
  ( 1 +  {r}^{2} \rme^{- \rmi 2 \varphi} + 
  {r}^{4} \rme^{- \rmi 4  \varphi}  + \ldots ) \, ,  \nonumber 
\end{eqnarray}
where $ \varphi = \frac{2 \pi}{\lambda}  n d \cos \theta$
is the plate phase thickness. Here, $\lambda$ is the wavelength in vacuum
and $\theta$ the angle of refraction in the medium $n$, which is
related to the angle of incidence according to Snell's law.

The complex reflection and transmission coefficients (i.e., the ratios
$ A_{\mathrm{ref}} / A_{\mathrm{in}}$ and $ A_{\mathrm{trans}} / A_{\mathrm{in}}$,
respectively) are obtained by adding all the 
waves in (\ref{eq:TampFP1}); the result is
\begin{equation}
  \label{eq:TampFP2}
 R(\varphi) = r^\prime +
 \frac{tt^\prime r \exp(-\rmi 2 \varphi)}{1-r^2 \exp(-\rmi 2 \varphi)}
 \, ,
\qquad 
 T (\varphi ) =\frac{t t^{\prime} \exp( - \rmi \varphi )}
  {1- {r}^{2} \exp(-2 \rmi \varphi)} \,  .
\end{equation}
The quantities $r, r^{\prime}, t, t^{\prime}$ are given by Fresnel
formulas. For the typical operation parameters of a a realistic FP, the
dependence of these coefficients with the angle of incidence or the
wavelength can be neglected and take them as constants. We shall make
use of the  Stokes relations~\cite{Born:1999yq}
\begin{equation}
  \label{eq:Stokes}
  r^\prime = - r \, , 
\qquad 
  t t^{\prime} +  r^{2} = 1 \,  ,
\end{equation}
which constraints the reflectivity and transmissivity of both
surfaces. Equations~(\ref{eq:TampFP2}) can then be recast as
\begin{equation}
  \label{eq:TampFP}
  R ( \varphi ) = \frac{r [ \exp(-\rmi 2 \varphi) - 1]} 
  {1-r^2 \exp(-\rmi 2 \varphi)} \, ,
  \qquad
  T (\varphi ) =\frac{(1 - r^{2}) \exp(-  \rmi \varphi )}
  {1- {r}^{2} \, \exp( - 2 \rmi \varphi)} \,  .
\end{equation}
These responses can be seen as parametric curves in the complex plane.
First of all, we observe that $R (\varphi)$ is a $\pi$-periodic function,
while $T(\varphi)$ is $2\pi$-periodic. This translates into the fact that
when $T (\varphi)$ completes a revolution, $R(\varphi) $ makes two turns.

Probably, the easiest way to
draw them is by introducing polar coordinates 
\begin{equation} 
  R = |R| \exp ( \rmi \rho ) \, , 
  \qquad
  T = |T| \exp ( \rmi \tau ) \, . 
\end{equation}
For reasons that will become apparent soon, we call
$\mathcal{R} = | R |^{2}$ and $\mathcal{T} = | T |^{2}$. With this
parametrization, equations~(\ref{eq:TampFP}) read as
\begin{equation}
  \label{eq:param}
  \mathcal{R}  =  4 a^{2} \cos^{2} \rho  \, , 
  \qquad 
  \mathcal{T}   =  1 -  4 a^{2} \sin^2 \tau  \, ,
\end{equation}
where $a = {r}/(1 + {r}^{2})$ is a real parameter verifying
$ 0 \le a^{2} \le 1/4$.

\begin{figure}
  \centering{\includegraphics[height=5cm]{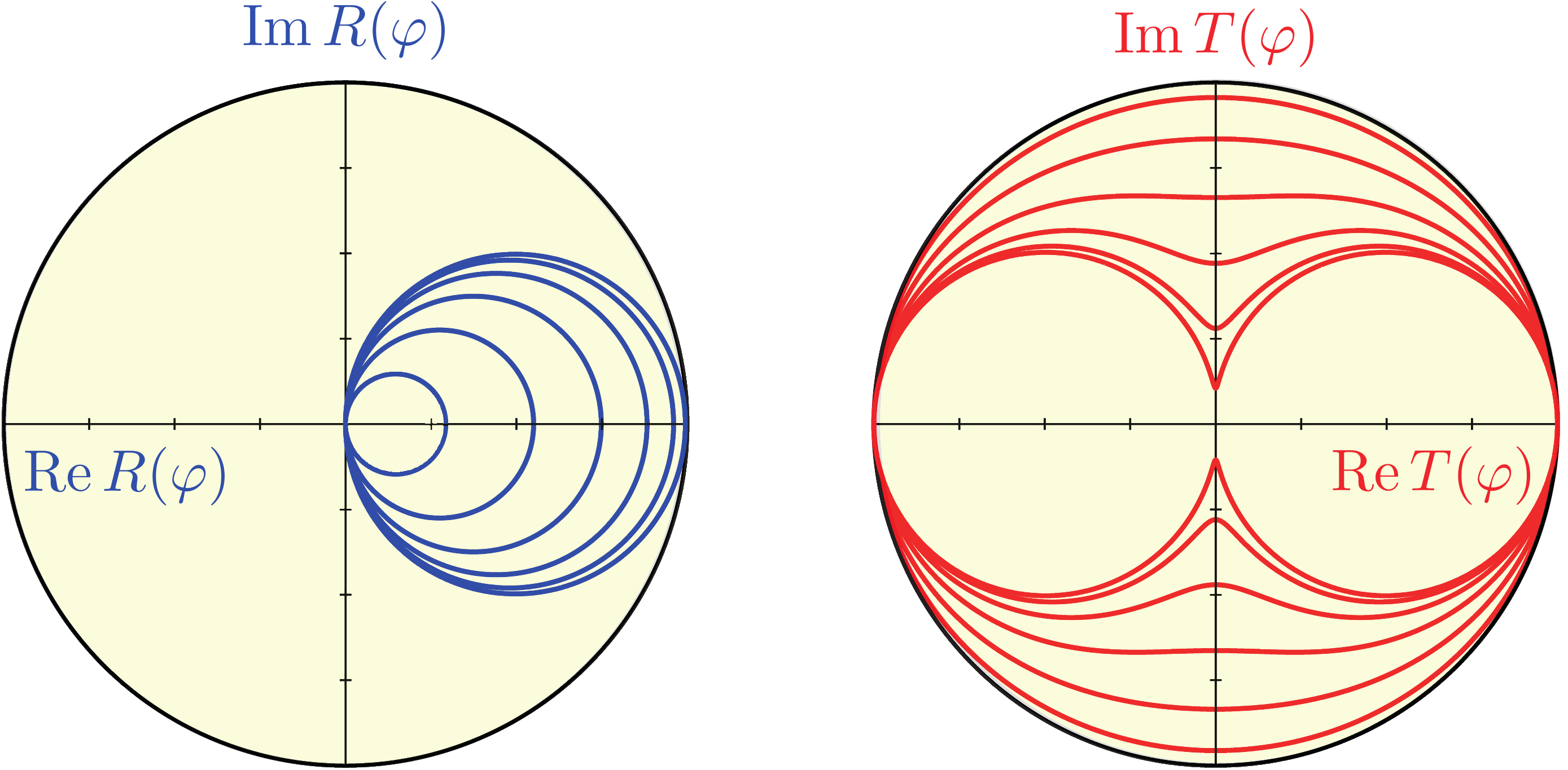}}
  \caption{Reflected (left) and transmitted (right) amplitudes by a
    FP.  The different curves correspond reflection coefficients
    ranging from $r = 0.15$ to $r = 0.90$ in steps of 0.15. For
    $R ( \varphi )$, the curves increase in size with $r$, while the
    converse happens for $T (\varphi)$.}
  \label{fig:RTImTvsphi}
\end{figure}

Both amplitudes lie inside the unit disk because one can immediately
check that
\begin{equation}
  \label{eq:ComRT}
  \mathcal{R} + \mathcal{T} = 1 \, , 
\end{equation}
which is just energy conservation.  The reflected amplitude
$R (\varphi) $ is readily identified as a circle of radius $a$
centered in the point $(a, 0)$ of the real axis. Actually, it can also
be expressed as $R=a[1+\exp(i2 \rho)]$, which confirms that.  The
transmitted amplitude $T (\varphi) $ describes a hippopede, a curve
full of stunning properties~\cite{Lawrence:1972aa, Shikin:1995aa}.
For $0 < a^{2} < 1/8$ it is an oval, and for $1/8 < a^{2} < 1/4$ it is
an indented oval, which tends to be an eight in the limit
$a^{2}=1/4$. ln figure~\ref{fig:RTImTvsphi} we represent $R(\varphi)$
and $T(\varphi)$ for several values of ${r}$, supporting these facts.

 %%%%%%%%%%%%%%%%%%%%%%%%%%%%%%%%%%%%%%%%%%%%%%%%%%
%%%%%%%%%%%%%%%%%%%%%%%%%%%%%%%%%%%%%%%%%%%%%%%%%%

The reflected amplitude passes through the origin for any value of
$r$: $R(\varphi)$ is zero for $\varphi=0$ and $\pi$ and traces the
circle clockwise, getting its maximum at $\varphi = \pm \pi/2$, where
$\rho = 0$ and then, according equation~(\ref{eq:param}),
$\mathcal{R}_{\mathrm{max}} = 4 a^{2}$.

On the other hand, the transmitted amplitude also describes the
hippopede clockwise. At $\varphi = 0$ and $\pi$,
$\mathcal{T} (\varphi)$ reaches its maxima, which are in the real axis
at $T =1$ and $-1$, respectively.  The minimum occurs at
$\varphi = \pi/2$ and $3 \pi/2$, where $\tau = -\pi/2$ and $-3 \pi/2$,
respectively.  Therefore $ \mathcal{T}_{\mathrm{min}} = 1 - 4a^{2}$,
which corresponds to half the waist of the hipoppede at its
indentation~\cite{Monzon:2015aa}.  Note that $a^{2} = 1/8$ happens for
$r= \sqrt{3 - \sqrt{8}} \simeq 0.4142$ and then
$\mathcal{R}_{\mathrm{max}} = \mathcal{T}_{\mathrm{min}} = 1/2$.

We also observe that the quotient
\begin{equation}
  \frac{R(\varphi)}{T (\varphi)}=   
  \rmi \, \frac{2 r} {1- r^{2}}  \sin \varphi 
\end{equation}
is an imaginary number. Therefore for a transparent symmetric system,
as the one we are dealing with, we have
\begin{equation}
  \label{eq:tau-rho}
  \rho (\varphi) - \tau (\varphi)  = \pm \frac{\pi}{2} \, ,
\end{equation}
so the reflected and transmitted amplitudes are always at quadrature.

To round up this discussion, we examine the local slopes
$\dot{\rho}(\varphi)$ and $\dot{\tau} (\varphi)$, the dot denoting
derivative respect to the parameter.  They are just  the rates at which
the curves are traced out. Indeed, they entail a valuable physical
interpretation. If we focus for simplicity at $\tau$, we can write
\begin{equation}
  \label{eq:2}
  \frac{\dot{\tau}}{\mathcal{T}} = 
\frac{\tan \tau}{\tan \varphi} =
  - \frac{1 + {r}^{2}}{1 - {r}^{2}}  \, ,
\end{equation}
the negative sign reflecting that the curve is oriented clockwise.
This quotient is thus independent of $\varphi$: when the transmitted
amplitude is large, so is the velocity and the opposite.

Moreover, $\dot{\tau} (\varphi)$ admits a crystal clear
interpretation.  Actually, a straight application of the chain rule
gives
\begin{equation}
  \label{eq:dpm}
  \mathfrak{t}_{g} = \dot{\tau} (\varphi) \,  \mathfrak{t}_{c} \, , 
\end{equation}
where the group- and phase-delay times are,
respectively,~\cite{Yu:2001aa}
\begin{equation}
  \label{eq:dpm2}
  \mathfrak{t}_{g} = \frac{\rmd\tau}{\rmd \omega}   \, ,   
  \qquad
  \mathfrak{t}_{c}= \frac{\rmd\varphi}{\rmd\omega} \, , 
\end{equation}
with $\omega$ the angular frequency.  The time $ \mathfrak{t}_{c}$ is
the single-pass time inside the plate (for a non-dispersive material,
this is $n d \cos \theta/c$, with $c$ the velocity of light in
vacuum), and $ \mathfrak{t}_{g}$ is a measure of the time needed for a
signal to propagate through the device. In a crude model,
$ \mathfrak{t}_{g}$ is the time a photon spends before it is
ultimately transmitted (which is also called the dwelling
time~\cite{Hauge:1989aa}). Hence, $ \dot{\tau} (\varphi)$ may be
viewed as an enhancement factor of the dwell due to the FP itself.

\subsection{Variations on the same subject}

\subsubsection{Phasors.---}

To facilitate the discussion, in what follows we shall restrict our
attention to the transmitted field.  One can alternatively interpret
$A_{\mathrm{trans}}$ in equation~(\ref{eq:TampFP1}) as a sum of
phasors (or rotating vectors), each one representing the field
transmitted into the plate in a round trip. The length of every
individual phasor is just the modulus of the corresponding term in the
sum, whereas its orientation is the phase of such a term.

%%%%%%%%%%%%%%%%%%%%%%%%%%%%%%%%%%%%%%%%%%%%%%%
\begin{figure}[b]
  \centering{\includegraphics[height=5.3cm]{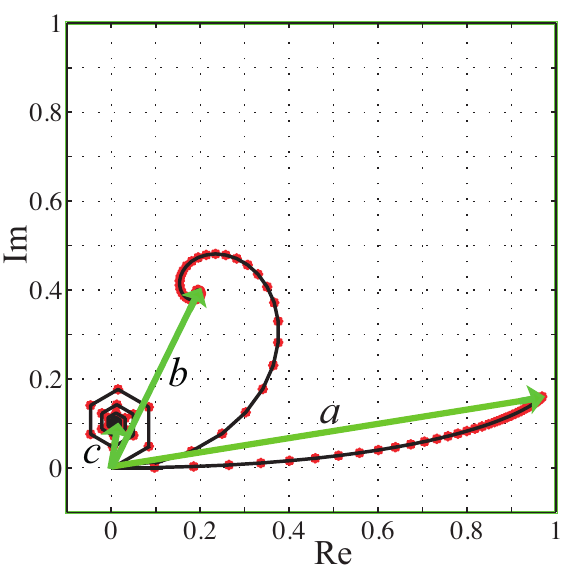}}
  \caption{Representation of the first fifty terms
   in equation~(\ref{eq:TampFP1}) as phasors, as well as the resulting 
  transmitted  amplitude for an FP interferometer. The arrows are
    indicated by red points and the total  field is plotted in green. 
    The horizontal axis represents signals in phase
    with the input field, and the vertical axis represents signals $\pi/2$
    out of phase. The three examples correspond to $r=0.95$ and  phase
    lags $\varphi$ of  (a) $\pi/360$,  (b) $\pi/30$ and (c) $ \pi/6$.} 
  \label{fig:phasor}
\end{figure}
%%%%%%%%%%%%%%%%%%%%%%%%%%%%%%%%%%%%%%%%%%%%%%%

Figure~\ref{fig:phasor} shows the phasors and the resulting
transmitted amplitude for near-resonant, off-resonant, and
intermediate conditions. Phasors in phase with the incident field are
plotted along the horizontal axis, while those shifted by $\pi/2$ are
plotted along the vertical axis.

Near resonance, all phasors acquire a phase shift close to an integer
multiple of $2\pi$ after each round trip: they are nearly parallel and
constructively interfere to yield a large transmitted amplitude.  In
off resonance, the components of the field acquire a wide range of
phase shifts after circulating within the plate and tend to
destructively interfere and produce a negligible overall field.

In the same vein, higher values of ${r}$ result in phasors that remain
in the FP for a longer time.  In this case, phasors can still be
of a significant amplitude when, after many round trips, they have
acquired enough phase to destructively interfere with the other
phasors in the FP.  Input fields whose frequencies are several
linewidths from the resonance condition can be considered to be
directly reflected, and the resulting field is small.

This approach highlights the need for exceptionally small tolerances
for the mirrors flatness.  The high transmission and narrow linewidth
occur because all the phasors become accurately aligned when
$\varphi =0$ or $\varphi = \pi$. But if there are small errors in the
phase because the reflections came from different parts of the
plate, the alignment is less accurate and the linewidth increases.

Phasor analysis is a useful mathematical tool for solving problems
involving linear systems in which the excitation is a periodic time
function~\cite{Feynman:2011aa}.  Although the reader most likely has
encountered this phasor notation in a variety of areas, its use
in the context of FP is rare~\cite{Siegman:1986aa}, even if it allows
to treat situations for which the standard method, as developed in the
previous subsection, fails: for example, the temporal response of a FP
to changes in cavity length and frequency of the incident
field~\cite{Lawrence:1999ba}.

\subsubsection{Equivalent diffraction grating.---}
 \label{sec:equiv}

\begin{figure}[b]
  \centering{\includegraphics[height=5.5cm]{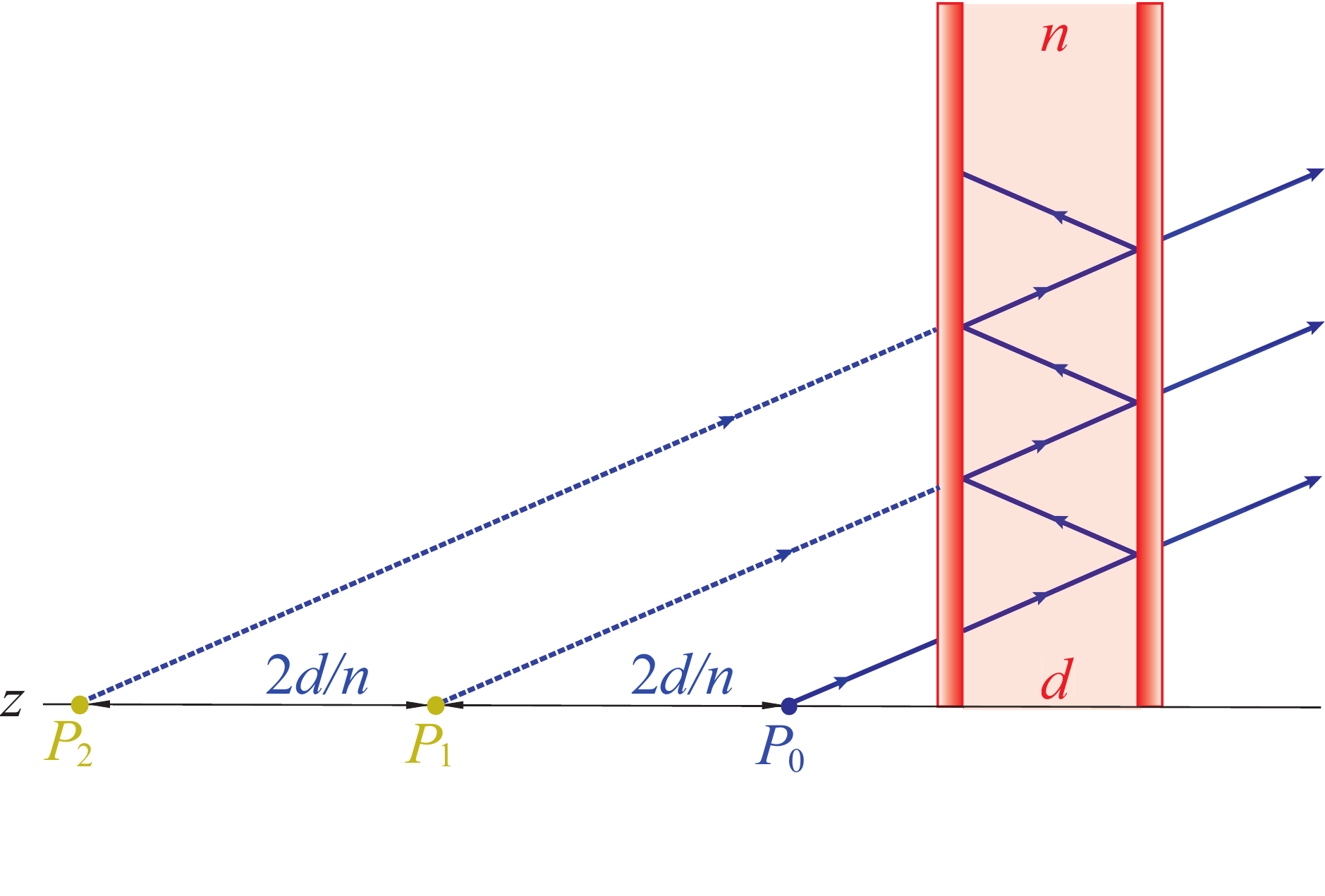}}
  \caption{FP illuminated by an ideal point source $P_{0}$ and the
    series of virtual sources $P_{1}$, $P_{2}, \ldots$ that represent
    the action of the transmitted waves.}
  \label{fig:equivS}
\end{figure}

 Let us think for a moment of the FP as illuminated by spherical
 wavefronts arising from an ideal point source $P_{0}$, as  
 sketched in figure~\ref{fig:equivS}.  The action of the FP thus appears as
 the interference of the wavefronts emanating from $P_{0}$ and the
 virtual sources $P_{1}$, $P_{2}, \ldots$, separated by a distance
 $2d/n$ (we assume $n^{\prime} = 1$), and amplitudes decreasing as in
 equation~(\ref{eq:TampFP1}). This simplified
 picture holds only within the limits of the paraxial approximation;
 otherwise, aberrations appear, although this plays no major role for
 our discussion here.

%%%%%%%%%%%%%%%%%%%%%%%%%%%%%%%%%%%%%%%%%%%%%%%%
%%%%%%%%%%%%%%%%%%%%%%%%%%%%%%%%%%%%%%%%%%%%%%%%

This opens a new way of looking into multiple-beam interference: the
beams being interfered can be though of as originating from equally
spaced coherent point sources in a line, which bears a close
resemblance to a diffraction grating. Actually, if we assume a
``shaded'' diffraction grating, with a pupil function of
$\exp(- \eta z)$, with $z$ being the position along the axis
perpendicular to the mirrors, one recovers exactly the amplitudes in
(\ref{eq:TampFP1}) once we identify
$\eta = n \ln (1/r^{2})/2d$~\cite{Brooker:2003aa}, as we shall confirm
soon.

It is worth observing that the separation between images,
$2d/n$,  is a bit of a surprise, since one expects $d$ to be multiplied
and not divided by $n$.  But if we use the diffraction grating
equation $m \lambda  = (2d/n) \cos \theta^{\prime}$ (defining
$\theta^{\prime}$ as the angle measured from the plane of the
grating), everything works out since $\cos \theta^{\prime} \simeq 1 -
\case{1}{2} \theta^{\prime 2} \simeq \case{1}{2} n^{2} \theta^{2}$.

%%%%%%%%%%%%%%%%%%%%%%%%%%%%%%%%%%%%%%%%%%%%%%%%
\begin{figure}
  \centering{\includegraphics[height=5cm]{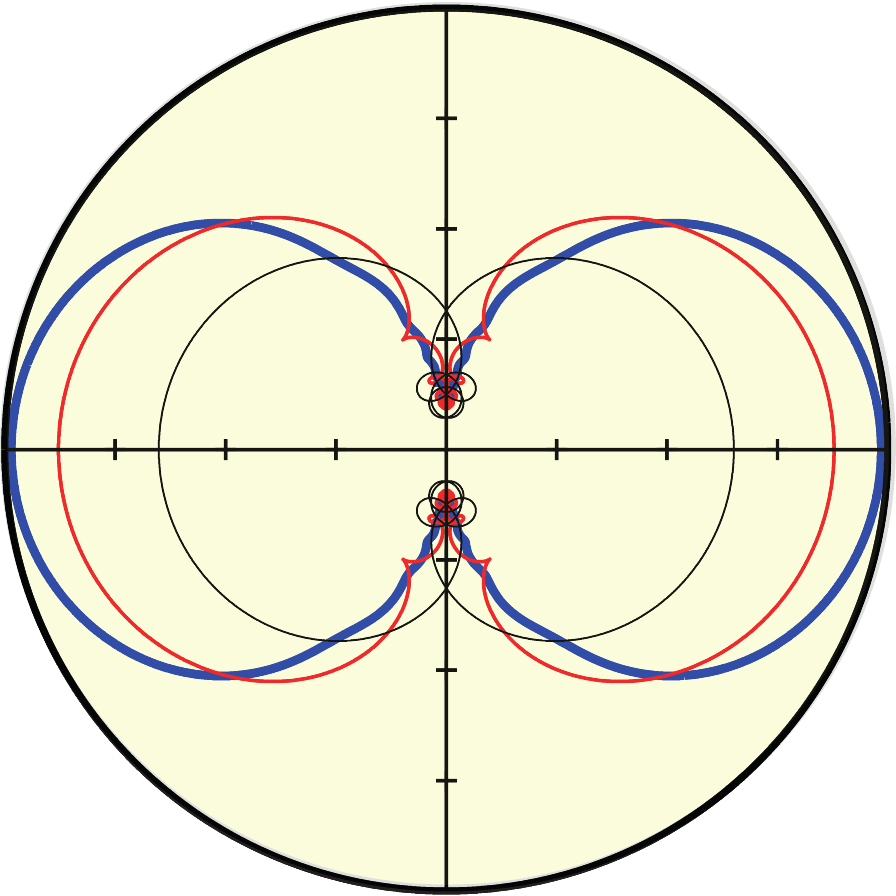}}
  \caption{Transmitted amplitude for a transparent FP with plates
    coefficient ${r} = 0.90$. We have pondered a finite number of terms
    in (\ref{eq:TampFP1}): 5 for the black curve, 10 for the red and
    20 for the blue. For this last case, the difference with the ideal
    hippopede is negligible.}
  \label{fig:preHyp}
\end{figure}
%%%%%%%%%%%%%%%%%%%%%%%%%%%%%%%%%%%%%%%%%%%%%%%%

Our physical intuition suggests that, since these secondary sources get
fainter farther away from the real source, only an effective number of
them contribute. This can be equivalently formulated as how many terms
must be retained in the series (\ref{eq:TampFP1}) for it to be
accurate enough. In figure~\ref{fig:preHyp} we have represented the
amplitude resulting when one takes into account 
5, 10 and 20 terms. The convergence to the ideal curve,
 shown in figure~\ref{fig:RTImTvsphi}, is pretty fast, although it
 depends on the values of $r$. A more quantitative discussion of this
question will be presented in section~\ref{sec:quadis}.

Another practical aspect of this construction is to give
intuition about what happens when the planes are not quite
parallel. The points $P_{0}, P_{1}, P_{2}, \ldots$ are now equally
spaced around a circle and the resulting diffraction grating then
gives a high-order Bessel function~\cite{Mcgloin:2005aa}, which has
associated oscillations on one side of the peak, instead of the Airy
function.  The effective number of reflections is now replaced by the
number of equivalent sources that lie in half the circle (the greatest
physical extent), and this gets smaller as the angle between the
mirrors increases.

From an electric-engineering perspective, these secondary sources 
constitute a far-field uniform linear array antenna with excitation
amplitudes arranged according to a geometric sequence.  Such a
configuration can be addressed with the standard methods of antenna
theory~\cite{Balanis:2016aa}. The high gain and high directionality
of this setup appear from the amazing properties of the
FP. 

\subsubsection{Self-consistent field.---}
\label{sec:FPres}

\begin{figure}[t]
  \centering{\includegraphics[height=5cm]{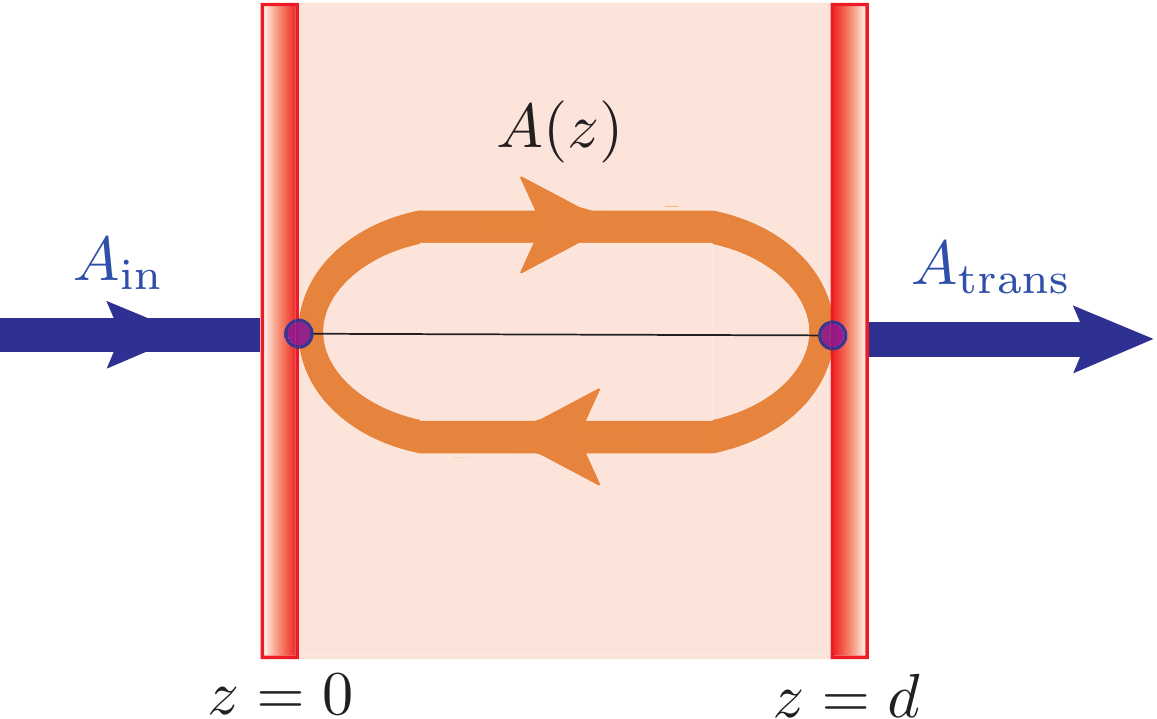}}
  \caption{Schematic block diagram of an FP. We indicate the
    intracavity field, as well as the input and transmitted ones (the
    reflected is omitted for clarity). The axis $z$ is perpendicular
    to the mirrors.}
  \label{fig:FPIntra}
\end{figure}

Instead of looking at the field inside the plate as arising from
multiple reflected waves, we can just think of it as an intracavity
field which has to travel the length of the cavity twice in opposite
directions to complete a round trip~\cite{Macleod:2010aa}.  We take as
before the $z$ axis perpendicular to the mirrors, and denote by $A(z)$
the complex amplitude of this field at an arbitrary plane $z$ inside
the mirrors, as schematized in figure~\ref{fig:FPIntra}.  The
self-consistency condition can be concisely formulated as
\begin{equation}
  A (0) =   A (0) \, {r}^{2} \exp(- \rmi 2 \varphi ) + 
  {t}^{\prime} \, A_{\mathrm{in}}  \, .
\end{equation}
This amount to imposing that the wave circulating after the first
mirror has to be reconstructed by interference of the internal wave
after one full round trip and the incident
wave~\cite{Meschede:2004aa}. In other words, we impose that  
the field distribution reproduces itself after one round trip in the
cavity.  From here, we have
\begin{equation}
  A (0) = \frac{t^{\prime}}{1 - {r}^{2}  \exp(-\rmi 2 \varphi )}
  A_{\mathrm{in}}  \, ,
\end{equation}
and if we take into account that
$A (d ) = A (0) \exp(- \rmi \varphi )$ and the boundary condition
$ A_{\mathrm{trans}} = {t} \, A (d)$, we immediately get that the
transmission coefficient $ T = A_{\mathrm{trans}}/A_{\mathrm{in}}$
coincides with (\ref{eq:TampFP}), obtained as a summation of infinite
waves. 

%%%%%%%%%%%%%%%%%%%%%%%%%%%%%%%%%%%%%%%%%%%%%%%%
%%%%%%%%%%%%%%%%%%%%%%%%%%%%%%%%%%%%%%%%%%%%%%%%

We stress the leading role played by the boundary conditions in this
idea. The method is at the realm of the transfer matrix
approach~\cite{Sanchez-Soto:2012aa}, which is of great value in
dealing with multilayered structures.

\begin{figure}[b]
  \centering{\includegraphics[height=2.75cm]{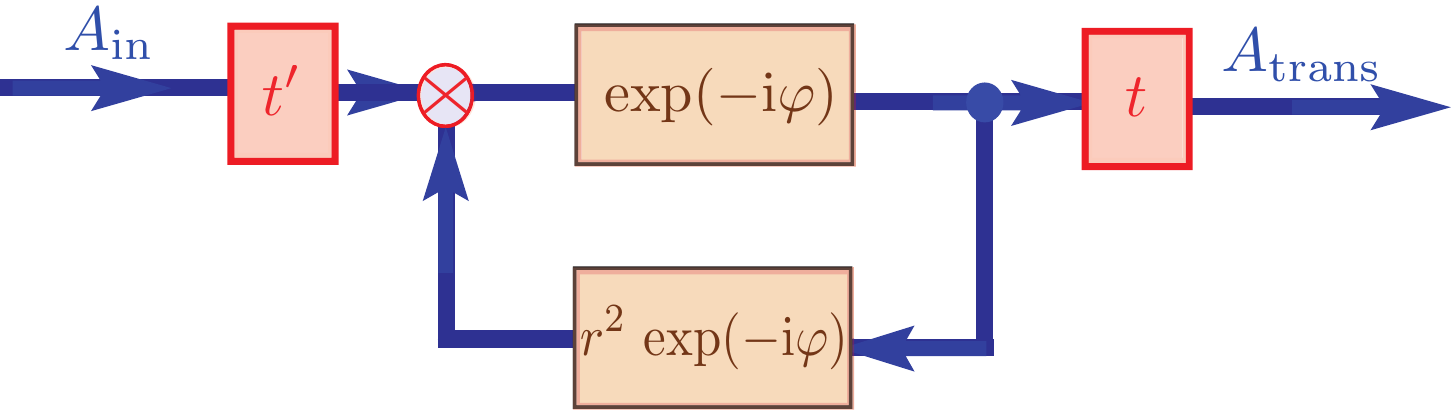}}
  \caption{Structure diagram of the FP interferometer in figure 1.}
  \label{fig:FPLS}
\end{figure}

These results can also be reinterpreted using elementary notions of
the theory of linear systems~\cite{Chen:2012aa}.  Indeed, the FP can
be envisioned as a closed-loop system~\cite{Donges:1997aa}, as
outlined in figure~\ref{fig:FPLS}. Apart from the mirrors, we have a
main box, which is just a delay line representing the propagation of
the intracavity field.  We have as well a feedback path with a
frequency response ${r}^{2} \exp(- \rmi \varphi )$. A direct
application of the standard rules leads to the same transmission
coefficient $T (\varphi )$.  Note that $T (\varphi )$ appears here as
the transfer function and accordingly its poles and
zeros provide full information of the response of the system. From
this perspective, the FP is a linear, time-invariant and stable
system.

The self-consistency condition can be worked out in the time domain,
instead of the frequency domain. The intracavity field at an arbitrary
time $\mathfrak{t}$ is determined by the condition~\cite{Rakhmanov:2002aa}
\begin{equation}
  A(\mathfrak{t}) = A (\mathfrak{t} - 2 \mathfrak{t}_{c} ) \, {r}^{2} \exp(- \rmi 2 \varphi) 
  + t^{\prime} \, A_{\mathrm{in}}  \, ,
\end{equation}
where the single-pass time $\mathfrak{t}_{c}$ is given in (\ref{eq:dpm2}).
Taking the Fourier transform on both sides of this equation yields the
basic cavity response function as before.  This formulation is
especially suitable for cavities with moving mirrors, which is the
basic scheme for the recent detection of gravitational
waves~\cite{Abbott:2016aa}.

 %%%%%%%%%%%%%%%%%%%%%%%%%%%%%%%%%%%%%%%%%%%%%%%%%
%%%%%%%%%%%%%%%%%%%%%%%%%%%%%%%%%%%%%%%%%%%%%%%%% 

\subsubsection{Impulse response.---}
\label{sec:nhf}

The previous example suggests to examine the general FP response in
the time domain~\cite{Kafri:1973aa}. In this way, we can deal with dynamical
situations, such as, e.g., pulse propagation in a FP. To this end, we
apply a short pulse whose amplitude can be approximated as
$A_{\mathrm{in}} \delta (\mathfrak{t})$, where $\delta$ stands for the Dirac
delta function. Neglecting any material dispersion, the transmitted
amplitude, for $\mathfrak{t} > 0$, is
\begin{equation}
  \label{eq:train1}
A_{\mathrm{trans}} ( \mathfrak{t} ) = A_{\mathrm{in}} \, t t^{\prime} 
\sum_{m=0}^{\infty} 
\delta ( \mathfrak{t} - \mathfrak{t}_{c}  - 2m \mathfrak{t}_{c} )  
\exp [ - \gamma_{c} ( \mathfrak{t} - \mathfrak{t}_{c} ) ]  \, , 
\end{equation}
and we have defined $\exp( - \gamma_{c} \mathfrak{t}_{c} ) = r^{2}$.  This
$A_{\mathrm{trans}} ( \mathfrak{t} )$ consists of a train of short pulses
separated in time by $2 \mathfrak{t}_{c}$, but in each round trip the pulse
amplitude is decreased by $r^{2}$. Rather than Fourier transforming
(\ref{eq:train1}) and summing up the resulting series, we follow an
alternative route and rewrite (\ref{eq:train1}) as
\begin{equation}
  \label{eq:train2}
  A_{\mathrm{trans}} ( \mathfrak{t} ) = A_{\mathrm{in}} \, \frac{t t^{\prime}}{r^{2}} \
  \mathrm{comb}_{2 \mathfrak{t}_{c}}  (\mathfrak{t}) \, \exp ( - \gamma_{c}  \mathfrak{t} ) \,
  H (\mathfrak{t})  \, .
\end{equation} 
Here the Dirac comb is $\mathrm{comb}_{\mathfrak{t}_{c} }
(\mathfrak{t})  = \sum_{m=- \infty}^{\infty} \delta
(\mathfrak{t} - m \mathfrak{t}_{c})$
and $H (\mathfrak{t})$ is the Heaviside step function that ensures that the
exponential is only for positive times. Since the Fourier transform of
a Dirac comb is a Dirac comb, the convolution theorem, used backwards,
suggests that the frequency response should be a convolution of an
infinite comb of frequencies and a building block that is the
transform of the one-sided exponential decay: this is just
$1/ (\gamma_{c} + \rmi \omega)$.  Moreover, since the convolution with
a delta function $\delta (\mathfrak{t} -m \mathfrak{t}_{c})$ is equivalent to shifting
the function by $m \mathfrak{t}_{c}$, convolution with the Dirac comb
corresponds to replication or periodic summation. In consequence, we
get
\begin{equation}
  \label{eq:2isim}
  T (\varphi ) =   \sqrt{\frac{1 - {r}^{2}}{1 + {r}^{2}}} 
 \sum_{m=- \infty}^{\infty}
 \frac{1}{(\case{1}{2}  \Gamma)  + \rmi (\varphi-m \pi )} \, ,
\end{equation}
where we have used the adimensional variable $\varphi$ and
$\Gamma = \gamma_{c} \mathfrak{t}_{c} = \ln (1/r^{2})$ is the decay
$\gamma_{c}$ in adimensional units.  $ T (\varphi )$ thus appears as a
sum of infinity curves displaced by integer values of $\pi$.

\section{Intensity response of a Fabry-Perot}

\subsection{The Airy distribution}

\begin{figure}[t]
   \centering{\includegraphics[height=5.2cm]{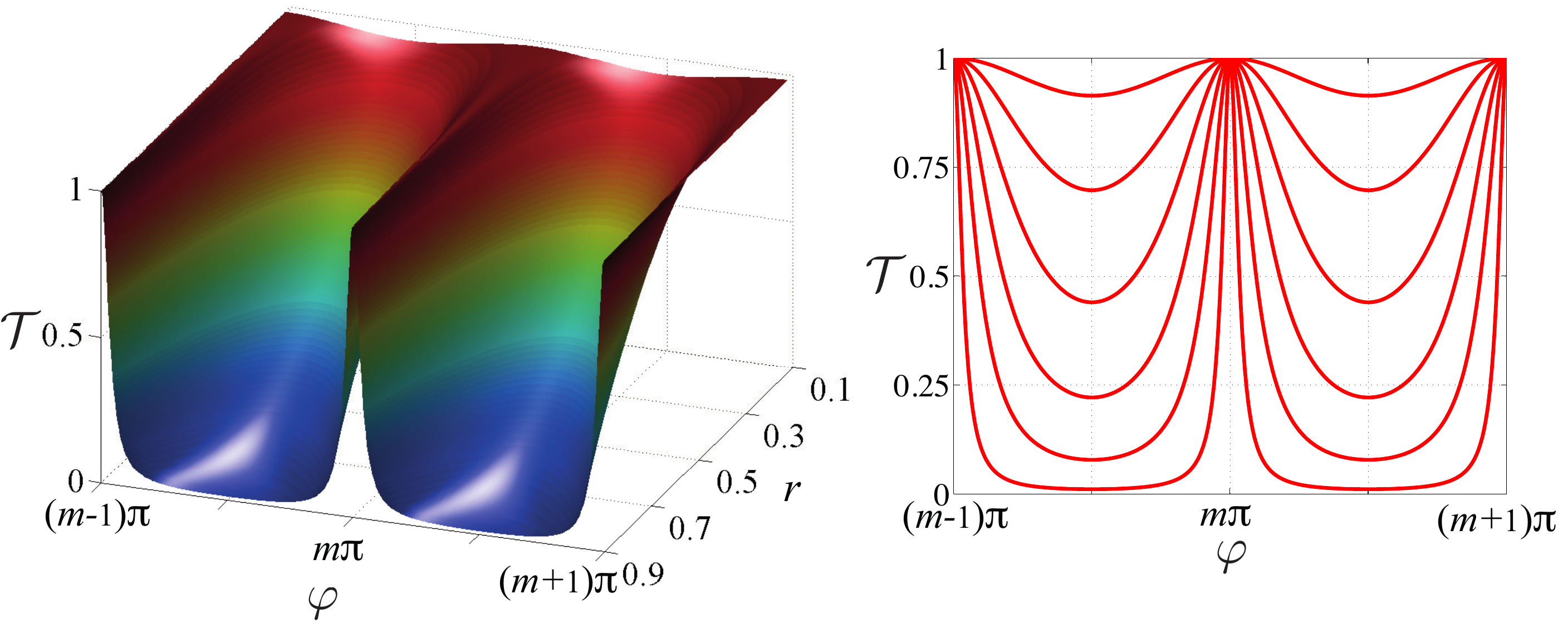}}
  \caption{(Left) Transmissivity $\mathcal{T}$ of the FP as a function of the
    phase shift $\varphi$ and the parameter $ r $. (Right) Cuts of the
  previous plot for reflection coefficients 
   ranging from $r = 0.15$ to $r = 0.90$ in steps of 0.15. }
  \label{fig:TphiF}
\end{figure}

The intensity response of the FP (i.e., the ratio of the transmitted
intensity to the incident one) is just $\mathcal{T} = | T |^{2}$, a
notation that was already anticipated in section~\ref{sec:mbi}. 
Consequently, it can be readily obtained from our previous discussion 
on the amplitudes. The final result is
\begin{equation}
  \label{eq:TintFP}
  \mathcal{T} (\varphi)  =  
  \frac{1}{1 + \mathfrak{f} \sin^{2} \varphi } \, ,
\end{equation} 
where the parameter $\mathfrak{f}$ is
\begin{equation}
  \mathfrak{f} = \frac{4 {r}^{2}} {(1- {r}^{2} )^2} \, .
\end{equation} 
Equation~(\ref{eq:TintFP}) is the time honored Airy distribution.  In
figure~\ref{fig:TphiF} we plot $\mathcal{T}$ as a function of the
phase thickness $\varphi$ and ${r}$. As ${r}$
increases, the minima of $\mathcal{T}$ fall and the maxima become
sharper. In the limit of high ${r}$, the pattern consists on
narrow bright fringes on an almost completely dark background.
Moreover, since we are not considering absorption in the plate, the
peak transmittance is unity (for any value of $r$).  

%%%%%%%%%%%%%%%%%%%%%%%%%%%%%%%%%%%%%%%%%%%%%
%%%%%%%%%%%%%%%%%%%%%%%%%%%%%%%%%%%%%%%%%%%%%

Consequently, one requires an increased reflectivity $r^{2}$ and this
is accomplished by coating the plates surfaces with a mirror. In what
follows, we assume that such a mirror is lossless.  In that case, the
Airy formula still holds provided we interpret ${r}$ as the reflection
coefficient of the mirror (which becomes, in general, a complex
number). This adds to the plate phase thickness $\varphi$ a phase
change on the reflection at the mirrors.  In general, both modulus and
phase of the complex ${r}$ depend on the angle of incidence and the
dispersion properties of the material, albeit such a variation can be
disregarded for most practical purposes.

The function $\mathcal{T} (\varphi)$ is $\pi$-periodic. The frequency
separation of adjacent fringes is called the free spectral range (FSR)
\begin{equation}
  \label{eq:FSR}
  \Delta \omega_{\mathrm{FSR}} = \frac{\pi c}{d} \, .
\end{equation}
The sharpness of the fringes is conveniently measured by their full width
at half maximum (FWHM), which is the frequency width 
between the two points on either side of a maximum where the intensity
falls to half its maximum value. When $\mathfrak{f}$ is sufficiently
large, this width is~\cite{Born:1999yq}
\begin{equation}
  \label{eq:FWHM}
  \Delta \omega_{\mathrm{FWHM}} = \frac{c}{d} 
  \frac{1- {r}^{2}}{r} =
  \frac{2c}{d} \frac{1}{\sqrt{\mathfrak{f}}} \, .
\end{equation}
The quotient between fringe separation and fringe width,
\begin{equation}
  \mathcal{N} = \frac{\Delta \omega_{\mathrm{FSR}}}
  {\Delta \omega_{\mathrm{FWHM}}}= 
\frac{\pi \sqrt{\mathfrak{f}}}{2} \, ,
\label{eq:defin}
\end{equation}
is known as the finesse.

%%%%%%%%%%%%%%%%%%%%%%%%%%%%%%%%%%%%%%%%%%%%%%%%%
\begin{figure}[t]
  \centering{\includegraphics[height=5cm]{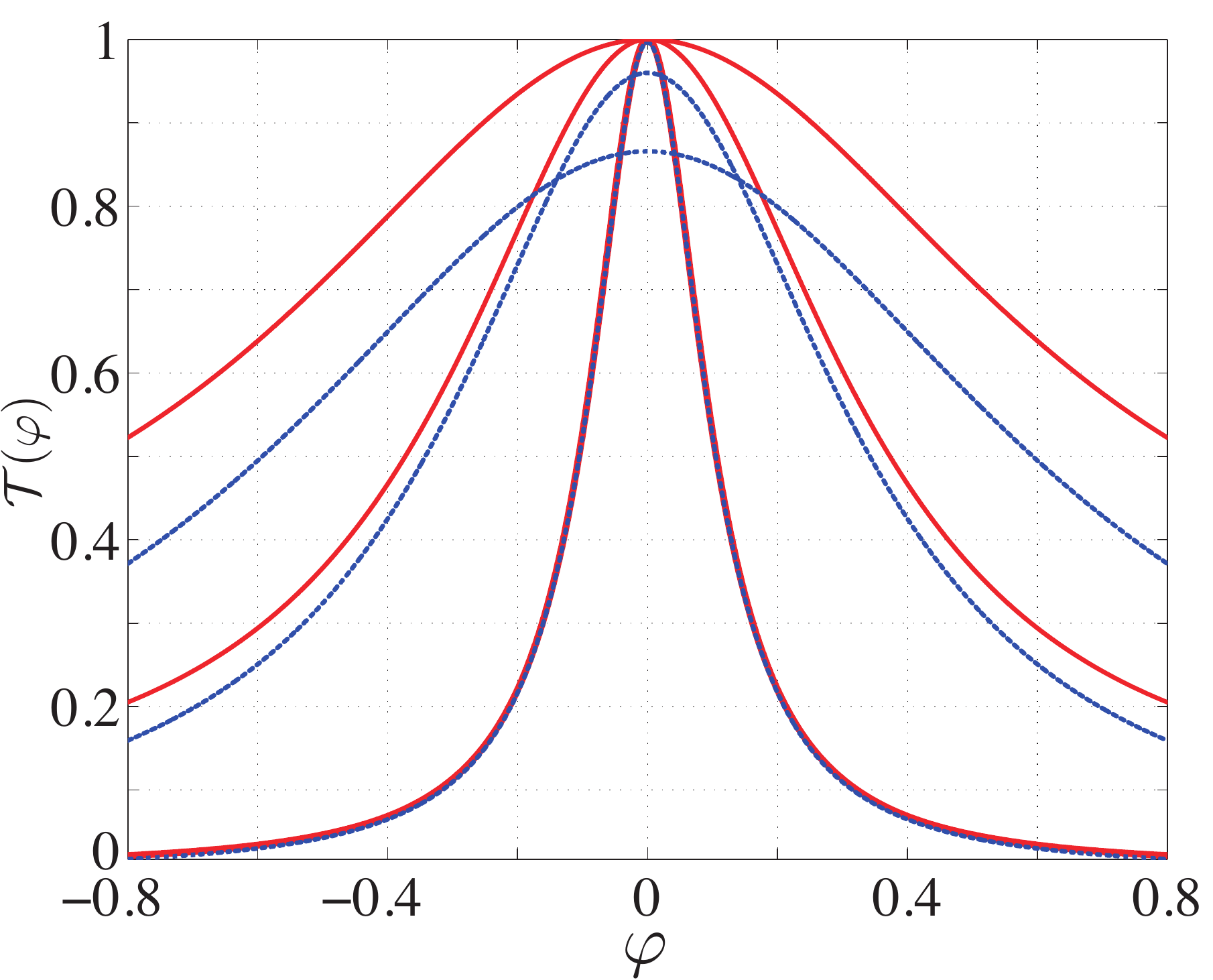}}
  \caption{Comparison between the Airy distribution (in continuous red
    line) and the Lorentzian approximation (in broken blue lines) for
    the values $r=0.5, 0.7$ and 0.9 (from top to the bottom).}
  \label{fig:ComANP}
\end{figure}
%%%%%%%%%%%%%%%%%%%%%%%%%%%%%%%%%%%%%%%%%%%%%%%%% 
 
Since $\Gamma = \ln (1/ {r}^{2})$,  $\mathcal{T} (\varphi)$ in
equation~(\ref{eq:TintFP}) can be recast as  
\begin{equation}
  \label{eq:Tnarrowpeak1}
  \mathcal{T} (\varphi) =  \frac{1 - {r}^{2}}{1 + {r}^{2}} 
 \left ( \frac{\sinh \Gamma}{\cosh \Gamma - \cos 2 \varphi} \right ) \, .
\end{equation} 
The term in parentheses is just a wrapped Lorentz
distribution~\cite{Mardia:2000aa}, so that   
\begin{equation}
  \label{eq:Tnarrowpeak2}
  \mathcal{T} (\varphi) =  
 \frac{1 - {r}^{2}}{1 + {r}^{2}} 
  \sum_{{m=-\infty}}^{\infty} 
  \frac{\frac{1}{2} \Gamma}
  {( \varphi -  m \pi)^{2} + (\frac{1}{2} \Gamma)^{2} } \, .
\end{equation} 
This is the intensity counterpart of equation~(\ref{eq:2isim}) and
could be obtained therefrom with a little of algebraic
effort~\cite{Koppelmann:1969aa}.  The Airy distribution appears in
this way as a sum of Lorentzian profiles of width $\Gamma$ displaced
by $\pi$ (which is just the FSR, in adimensional units).  If the width
$\Gamma$ is very small compared to the FSR, the contribution of every
Lorentzian far from its peak can be neglected. Consequently, around
the $m$th order, we can use
\begin{equation}
  \label{eq:Tnarrowpeak3}
  \mathcal{T}_{\mathrm{np}} (\varphi) \simeq  
  \frac{1 - {r}^{2}}{1 + {r}^{2}} 
  \frac{\frac{1}{2} \Gamma}
  { \varphi^{2} + (\frac{1}{2} \Gamma)^{2} }  \, ,
\end{equation}
so that each fringe has a Lorentzian shape. This is called the
narrow-peak approximation~\cite{Jacquinot:1960aa}, and  
now the finesse can be estimated as
\begin{equation}
  \label{eq:Finnp}
  \mathcal{N}_{\mathrm{np}} =  \frac{\pi}{\Gamma} = 
  \frac{\pi}{\ln ( 1/r^{2})} \simeq \frac{\pi}{1-r^{2}} \, ,
\end{equation}
which is equivalent to the expression (\ref{eq:defin}) in the limit
${r} \rightarrow 1$, which is precisely when this
approximation holds. This gives a better understanding of the
phenomenon of peak broadening when ${r}$ decreases.
In figure~\ref{fig:ComANP} we compare a single peak in the Airy
function with the corresponding narrow-peak approximation,
corroborating that it works well only for $r$ high enough.

\subsection{Resolving power}

The resolving power gives the ability to discriminate between
different wavelengths, to accomplish which it is first necessary to
define mathematically a separation criterion between two very close
maxima. Rayleigh criterion~\cite{Rayleigh:1879aa} is widely employed
for that purpose: it states that two intensity maxima are separated if
the maximum value of the first spot is superimposed on the first
minimum of the second spot~\cite{Born:1999yq}.  The resulting
resolution limit seems to be rather arbitrary and is based on
resolving capabilities of the human eye. Even more important, it
cannot be directly applied to the Airy intensity profile because of
the slow decrease of these values, so the minimum is located far from
the maximum~\cite{Juvells:2006aa}.

In our context, the most appropriate way of proceeding is perhaps to
adopt the Taylor criterion~\cite{Klein:1986aa, Fowles:1989aa} (also
called the FWHM criterion), which asserts that the separation of the
maxima is equal to the half-maximum width.  For the FP, it gives a
spectral resolving power (SRP)
\begin{equation}
  \label{eq:SRPdef}
  \mathrm{SRP} \equiv  
  \left | \frac{\lambda}{\Delta   \lambda_{\mathrm{FWHM}}} \right | =  
  \left | \frac{\omega}{\Delta \omega_{\mathrm{FWHM}}}  \right |   = 
  \frac{2d}{\lambda} \mathcal{N} = m \mathcal{N} \, ,
\end{equation}
where $m = 2d/\lambda$ is the interference order at normal
incidence. The SRP gives higher values for fringes near the centre of
the interferometric pattern and for values of $r \rightarrow 1$, which
means that closer wavelengths can be discriminated.

\subsection{Resolution revisited}

\subsubsection{$Q$ factor.---} 
When we consider the FP as an energy-storing system, a pivotal
parameter is the quality factor $Q$, defined for any damped
oscillator as
\begin{equation}
  \label{eq:Qfac1}
  Q = \omega \times \frac{\mathrm{energy \ stored}}
  {\mathrm{power \  dissipated}} \, .
\end{equation}
It is instructive to derive this expression for a FP and compare with
the SRP we have found before.

If the energy in the field is $W$, the factor $Q$ can be formulated as
\begin{equation}
  \label{eq:Qfac2}
  Q = \omega \frac{W}
  {W \displaystyle{\frac{c}{d}} \delta_{\mathrm{loss}}} = 
  \omega  \frac{d}{c} \frac{1}{ \delta_{\mathrm{loss}}} \, ,
\end{equation}
where $ \delta_{\mathrm{loss}}$ is the fractional loss per
transit. Since we are neglecting any diffractional effect, losses are
chiefly due to reflections in the mirrors and
$ \delta_{\mathrm{loss}} \simeq (1 -
{r}^{2})$~\cite{Siegman:1986aa}.
If we recall the definition of $\Delta \omega_{\mathrm{FWHM}}$, we
arrive at
\begin{equation}
  \label{eq:Qfac3}
  Q = \frac{2 \pi d}{\lambda} \frac{1}{1 - r^{2}} \, .
\end{equation}
This coincides with equation~(\ref{eq:SRPdef}) within a factor $r$,
which is accounted for by the small-loss approximation assumed
before. In consequence, as expected, the spectral resolving power is
synonymous with the quality factor. 

This comparison further establishes that if the energy losses are due
entirely to reflection losses at the mirrors, the resolving power may
be rigourously derived according to the multiple-beam treatment. When
losses due to diffraction become appreciable, the effective resolving
power must decrease, as $Q$ itself decreases.

This can be complemented with an easy uncertainty-principle argument:
a wave in the FP will have a decay time $\Delta \mathfrak{t}$ determined by
the fractional loss per transit  $ \delta_{\mathrm{loss}}$ and can be
roughly estimated as 
\begin{equation}
  \label{eq:Qalt}
  \Delta \mathfrak{t} = \frac{d}{c} \frac{1}{\delta_{\mathrm{loss}}} \, .
\end{equation}
For such a decaying wave, the Fourier-transform determined bandwidth
is
\begin{equation}
  \label{eq:QaltFT}
  \Delta \omega \simeq \frac{1}{\Delta \mathfrak{t}} \, ,
\end{equation}
which gives again the same value of $Q$.

\subsubsection{Equivalent sources.---}
\label{sec:quadis}
We recall that in \ref{sec:equiv} we have pinned down the FP action as
being equivalent to an infinite linear array of sources. However, we
anticipated that only an effective number of them contribute.  Indeed,
if we bear in mind that for a diffraction grating made of $N$
identical radiators one has~\cite{Born:1999yq}
\begin{equation}
  \label{eq:SRPdef2}
  \mathrm{SRP}_{\mathrm{grating}} = m N \, ,
\end{equation}
where $m$ is the order, this suggests, after a glance
at~(\ref{eq:SRPdef}), that the effective number of sources is
precisely the finesse $\mathcal{N}$.

This can be further confirmed from the time-domain approach in
\ref{sec:nhf}. For an incident short pulse, the FP transmitted train
of pulses fall in amplitude by a factor $1/\rme$ when
$r^{2N} =1/\rme$; that is, after
\begin{equation}
  \label{eq:rtri}
  N = \frac{1}{\ln (1/ r^{2})} \simeq \frac{1}{1-r^{2}}  = 
  \frac{\mathcal{N}_{\mathrm{np}}}{\pi}
\end{equation}
trips in the cavity.  Therefore, the number of effective round trips
is precisely the finesse (except for the factor $\pi$), and gives the
number of equivalent sources considered before.  The output pulse
train has a length (from the starting to the 1/e point) of
$n \, 2d \times N$, and we can loosely say that the coherence length after
traversing the FP has been increased by $N$~\cite{Brooker:2003aa}.

\section{Conclusions}

In summary, we have thoroughly explored several complementary
viewpoints on the FP response, both in amplitude and intensity.
Traditional discussions generally pinpoint the intensity, for
this is the variable measured in experiments. However, amplitude is
the natural arena to deal with the FP, as it is the variable for which
the superposition principle holds. Even though translating this
response into intensity is direct, the physical picture of the
approximations involved can be better appreciated using amplitude, as
we have illustrated in this work.

Yet very basic in nature, we hope that these results may be helpful in
updating the modern views on the operation of such a relevant setup.
We finally emphasize that the methods employed here are quite appealing
for they have branched into offshoots of importance for many other
modern physical theories.

\section*{Acknowledgments}
The original ideas in this paper have been developed and completed
with questions, suggestions, criticism, and advice from many students
and colleagues. Particular gratitude for help in various ways go to
G. Bj\"{o}rk, J. F. Cari\~{n}ena, H. de Guise, and A. B. Klimov. We
also thank two referees for their insightful comments that motivated
us to get deeper into the most delicate aspects of the problem treated
here. Financial support from the Spanish Ministerio de Econom\'{\i}a y
Competitividad (MINECO Grant FIS2015-67963-P) is gratefully
acknowledged.

\newpage
 
%\bibliographystyle{iopart-num}
%\bibliography{Fabry}

\providecommand{\newblock}{}

\end{document}